\documentclass[11pt]{article}
\usepackage[utf8]{inputenc}
\usepackage[T1]{fontenc}
\usepackage[margin=1in]{geometry}
\usepackage{graphicx}
\usepackage{booktabs}
\usepackage{microtype}
\usepackage[hidelinks]{hyperref}
\title{Memory as Infrastructure: Reliability Engineering for Persistent Agent Memory in Months-Long LLM-Assisted Development}
\author{Mike Helwig\\ \small Independent Researcher}
\date{August 2026}
\begin{document}
\maketitle
\begin{abstract}
LLM coding agents are crossing from task-scale to project-scale: single assisted efforts that run for months, across repeated context compactions, on codebases far larger than any context window. We report operational experience from one such effort (a research project under a single continuous Claude Code session line since January 2026, driving a 633,000-line codebase, with its memory subsystem continuously instrumented since July 2026) and describe SIx Harness, the open-source memory and continuity infrastructure that emerged from keeping it coherent. The harness combines per-project long-term memory (hybrid lexical--vector retrieval over SQLite, fully local), precision-gated context injection, anti-recurrence stores for decisions and dead-ends, conventions engineered to survive compaction, and, the part we argue is missing from operational practice, \emph{reliability engineering for the memory subsystem itself}: a session-start health gate with discriminated failure modes, heartbeat telemetry designed so that no enumerated failure mode can pass unrecorded, and alert-fatigue budgeting borrowed from SRE practice. We present what we believe is the first months-scale instrumented operational record of a persistent agent-memory subsystem in production development use: 78,933 hook invocations; 85 recorded failures, none silent: 84 in the subsystem's first three weeks, one since, none in the final 20 days; an injection layer whose ten-day precision instrument shows zero false fires against an intact denominator; and three production incidents traced from instrument reading to structural fix. From the record we distill seven design principles, state our limitations plainly (N=1, no control arm, self-reported), and publish a tagged pre-registered ablation protocol that any team can run with the released MIT-licensed kit.
\end{abstract}
\section{Introduction}

A working session with an LLM coding agent used to be an afternoon. Measured task horizons for AI agents have been doubling roughly every seven months [METR 2025], vendors now publish engineering guidance for agents that outlive their context windows [Anthropic 2025b], and serving systems treat context compaction as routine load-bearing infrastructure [ParallelCompaction 2026]. The regime this paper addresses is the far end of that curve: \textbf{one continuous assisted effort, one codebase, months of elapsed time, session boundaries beyond count}: the agent equivalent not of an employee's task but of an employee's tenure.

At tenure scale, the failures that matter most are not reasoning failures. They are memory failures, and we found they come in four kinds:

\begin{enumerate}\itemsep2pt
\item \textbf{Knowledge loss.} Context compaction summarizes away the connective tissue of long arcs. The agent resumes fluent but subtly unmoored: the plan survives, the \emph{why} does not.
\item \textbf{Repeated work.} The agent re-proposes what already exists. In our project's early record, three consecutive sessions each independently re-proposed a subsystem that was already built and sitting dormant, with the relevant design documents present in context each time.
\item \textbf{Context degradation.} Naive mitigation for (1) and (2), injecting more retrieved context, fails in the other direction. Injected blocks that are only sometimes relevant train the human operator (and plausibly the agent) to skim past them; once the channel is learned-noise, it is worse than absent. Irrelevant context also measurably degrades model output on its own [Shi 2023; Liu 2024; Chroma 2025].
\item \textbf{Silent memory death.} The meta-failure that makes the other three invisible: the memory system itself (an index, an embedding server, a hook pipeline) degrades or dies \emph{quietly}, while everyone keeps trusting it. Documentation research has long observed that artifacts go stale silently ("no crashes, no error messages") [OutdatedDocs 2023]; an agent-memory stack inherits exactly that property unless something watches it.
\end{enumerate}

Our central claim is that the fourth failure re-frames the other three: \textbf{persistent agent memory is infrastructure, and infrastructure needs reliability engineering}: health checks with actionable alarms, heartbeats, freshness objectives, and an economics of alerting that respects the operator's attention [SRE 2016; Breck 2017]. The 2026 literature has begun to argue that deployed agent memory degrades and deserves lifespan engineering [Aging 2026], and to taxonomize memory failure modes under stress [MemFail 2026], but on benchmarks, in vitro. What has been missing is an operational record: what actually breaks, what catches it, and what it costs, in real production use over months. This paper supplies one.

\textbf{Contributions.}

\begin{itemize}\itemsep2pt
\item \textbf{The system} (\S3): SIx Harness, an open-source (MIT) memory and continuity harness for Claude Code: local hybrid retrieval, precision-gated injection, anti-recurrence stores, compaction-surviving conventions, and a reliability layer for the memory subsystem itself.
\item \textbf{The record} (\S2, \S4): to our knowledge the first months-scale \emph{instrumented} operational record of a persistent agent-memory subsystem in production development use: 78,933 hook invocations, 85 recorded (zero silent) failures, an injection-precision record with an intact denominator, and three incident traces. The nearest neighbors are a one-month qualitative action-research study of 391 consecutive LLM sessions on one project [WrittenByAI 2026] and org-scale production GenAI telemetry that is memory-blind [WhatsCode 2026]; neither instruments the memory subsystem.
\item \textbf{The principles} (\S5): seven transferable design rules distilled from traced failures, including the observation that memory fails in \emph{two opposite directions} (rediscovering what exists and acting on decayed records), which demand deliberately opposed operational rules held in tension. The duality has recently been formalized as a benchmark metric [FAMA 2026], and its decayed pole is now benchmarked directly [STALE 2026]; we contribute the duality's workflow-level instantiation, its policy consequence, and in-vivo evidence.
\item \textbf{The invitation} (\S7): a pre-registered, publicly tagged ablation protocol over the released kit, so the causal questions our N=1 record cannot answer can be answered by anyone.
\end{itemize}

\section{The case: one session line, one codebase, eight months}

\begin{figure}[t]\centering
\includegraphics[width=\linewidth]{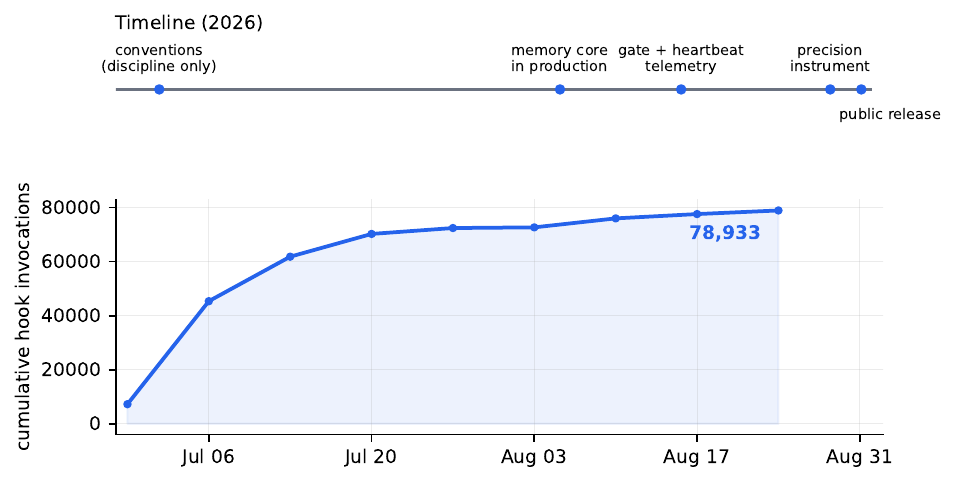}
\caption{The deployment timeline over the cumulative hook-invocation record (78{,}933 through 2026-08-30). Instrumentation begins with the gate and heartbeat layer on 2026-07-02.}\label{fig:timeline}
\end{figure}

The setting is a research codebase (a CUDA/C++ simulation engine with a large Python orchestration, verification, and tooling layer) under continuous development. As of this writing it comprises \textbf{633,446 lines across 1,936 code files}, plus a 448-file design corpus. Since January 2026, the eight months this paper covers, development has been conducted almost entirely through Claude Code, as \textbf{one logical session line}: a single conversational thread of work, re-anchored after every compaction. Session boundaries before instrumentation went uncounted; the instrumented final two months alone record \textbf{90 boundary crossings}, a rate implying a few hundred over the period. The human operator sets direction, makes rulings, and owns all approvals; the agent executes, and that includes, as discussed below, operating its own memory infrastructure.

\textbf{Timeline} (all anchors from git history and instrument stores; every number in this paper is re-derived by the documented queries in the artifact's numbers appendix):

\begin{itemize}\itemsep2pt
\item \textbf{January 2026.} The \emph{conventions} begin: a one-line-per-memory index file with a resume anchor read first after every compaction; first-class invariant files; append-only per-arc ledgers; a project constitution of standing rules. No tooling yet; discipline only.
\item \textbf{2026-05-24.} The \emph{memory core} enters production: \texttt{claude-mem}, a per-project long-term memory with hybrid BM25 + vector retrieval over a single SQLite file, ingesting docs, ledgers, transcripts, and git history.
\item \textbf{2026-07-02.} The \emph{reliability layer} enters production: the memory-health gate and per-hook heartbeat telemetry. From this date forward the subsystem's operational history is continuously instrumented: 78,933 hook invocations recorded through 2026-08-30.
\item \textbf{2026-08-19 $\to$ 08-28.} The \emph{precision instrument} is deployed: denominator-complete injection telemetry, recording every prompt, injection, latency, and gate verdict for ten calendar days. The window is defined by the instrument's deployment date, not selected after the fact.
\item \textbf{2026-08-29.} The harness is extracted, de-identified, and released as SIx Harness (MIT).
\end{itemize}

\textbf{Accumulated memory state} at time of writing: 19,065 indexed chunks; 389 curated memory files; 154 captured decisions and 186 captured dead-ends, of which 59 and 47 respectively survived operator triage into the confirmed anti-recurrence record; the rejects are not waste but the curation working (\S3.2).

\textbf{A note on who operates the system.} The memory infrastructure described here is not only \emph{for} the agent; it is substantially \emph{built and daily-operated by} the agent it serves, under the human operator's direction and review. The agent runs the triage queues, reads the gate line at session start, re-extracts stale corpora, and traces its own hook failures. We flag this for two reasons. First, it previews an operating mode we expect to become common: at tenure scale, an agent that cannot participate in maintaining its own continuity leaves that maintenance to whoever happens to be present at each session boundary. Second, it is a bias disclosure: the record is reported by the system's own operator-builders, human and agent alike (\S6).

\section{The system}

\begin{figure}[t]\centering
\includegraphics[width=\linewidth]{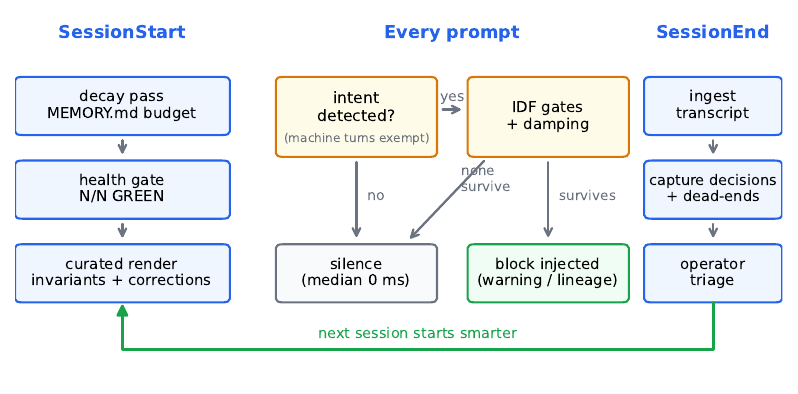}
\caption{The session lifecycle: SessionStart maintenance, gate, and render; the per-prompt gating cascade whose default is silence; the SessionEnd capture and triage loop that feeds the next session.}\label{fig:lifecycle}
\end{figure}

SIx Harness has five parts: a memory core, anti-recurrence stores, a precision-gated injection layer, a reliability layer, and a set of conventions shipped as templates. A session flows through them as follows: at SessionStart, a decay pass maintains the memory index file, the health gate probes the whole stack and prints one verdict line, and a curated render injects invariants, recent corrections, and recent decisions. On every prompt, an intent detector decides whether any retrieval work happens at all; if it does, relevance gates decide whether anything is injected. At SessionEnd, the transcript is ingested and candidate decisions and dead-ends are extracted for later triage.

\subsection{The memory core: boring storage, honest degradation}

The core is deliberately unexciting: one SQLite file per project, holding an FTS5 (BM25) index and a \texttt{sqlite-vec} vector table, populated by ingesters for docs, memory files, append-only ledgers, session transcripts, and git history, with watermark-based incremental refresh. Embeddings and capture-synthesis run against a local Ollama server; the system uses no cloud services and holds no API keys.

Two contracts matter more than the storage choice. First, \textbf{honest degradation}: if the embedding leg fails, search silently degrades to BM25-only rather than ever blocking a hook, and the failure is written to a log-once degradation store so the degraded period is visible afterwards (a design that \S4's incidents vindicate). Second, a \textbf{dimension guard}: vectors whose width does not match the index are refused at the door, never stored, which converts a class of misconfiguration from silent corruption into a detectable fault (\S3.4 closes the loop by detecting it at configuration time).

We explicitly did not adopt a graph store, an OS-metaphor memory manager [MemGPT 2023], or a temporal knowledge graph [Zep 2025]; the most graph-like structure in the system is the decision$\to$thread lineage links in plain SQLite. This is a considered position, not a gap: at single-project scale, our failures were never "the data model was insufficiently expressive" and always "some part of the pipeline was quietly wrong." Complexity spent on representation is complexity unavailable for reliability. (Structural code queries are delegated to an external graph tool the gate merely \emph{watches}; see \S3.4.)

\subsection{Anti-recurrence stores: curated, not scraped}

The repeated-work failure (\S1) is fought with records of what has been decided and what has been tried and killed. At SessionEnd, an extraction pass proposes candidate \emph{decisions} and \emph{dead-ends} from the transcript; they accumulate in a pending queue; the operator (in our deployment, usually the agent, batch-reviewed by the human) triages them (confirm, reject, retitle) before they enter the confirmed record. Of 154 captured decisions and 186 dead-ends, 59 and 47 survived triage. The reject rate is the point: an anti-recurrence store that ingests everything becomes the noise channel of \S1's third failure. Recent admission-control work reaches the same conclusion from the write side: gate what enters memory, argued on auditability grounds [AMAC 2026]; ours is the operator-triaged instantiation. Corrections (instances of the operator saying "no, do it this way") are additionally extracted as first-class records and re-surfaced at session start, rotation-aware so a fixed handful never monopolizes the channel.

The motivating incident predates these stores and is worth stating plainly. Early in the project, \textbf{three consecutive sessions each re-proposed building a subsystem that already existed} in the codebase in dormant form, with the design documentation injected into context each time. Perfect recall did not prevent rediscovery; the information was present and unused (a failure sub-mode that newer benchmarks now isolate as "retrieved-but-unused" [Mem2Act 2026]). The pattern stopped after two changes made in response: an inventory document the constitution requires consulting before proposing new infrastructure, and the DO-NOT-REBUILD injection of \S3.3. With both landing together amid ordinary project evolution, though, we cannot apportion credit between them; \S7's protocol exists to make such attributions testable.

\subsection{Precision-gated injection: the economics of the reader's attention}

The injection layer answers one question: \emph{when is it worth interrupting?} Its pipeline is a cascade of refusals, cheapest first. \textbf{Intent gates}: only prompts exhibiting build, investigate, or decision intent trigger any retrieval at all; on other turns the hook does no work. \textbf{IDF relevance gates}: a candidate injection must be supported by rare-token overlap between prompt and item (summed inverse-document-frequency above threshold, minimum word-boundary matches), so generic vocabulary cannot trigger it. \textbf{Per-session damping} caps repeats. A \textbf{system-turn exemption} excludes machine-generated turns: agent-completion callbacks and similar mechanical events that flow through the same prompt hook but present no decision point to any reader (\S4.3). What passes all gates is injected as one of three block types: a DO-NOT-REBUILD warning naming the existing subsystem, a stale-claim warning, or decision lineage.

A word on who the reader is. In this deployment the injected block is read by whoever authors the next action: the human when steering, the agent when executing a human-authored instruction. Both readers are subject to the same channel economics: a block that fires irrelevantly spends trust that only relevant firings restore. Before the relevance gates, the operations log records the stale-claim block firing on \textbf{roughly four in five intent-bearing prompts}: topically adjacent often enough to be defensible and useful rarely enough to be skimmed. (This contemporaneous measurement's full sample denominator was not preserved; we report it as context, not as a comparison leg; the instrumented window below stands on its own denominator.) Model-side evidence that irrelevant context degrades output [Shi 2023; Liu 2024; Chroma 2025] argues for injecting less; the reader-side skim reflex argues for \emph{silence as the default}. Our gate is the corpus-statistics analog of the learned selective-retrieval line [Self-RAG 2024; FLARE 2023; TARG 2025], chosen over learned gating for zero added latency and full auditability: every gate decision is reconstructable from the stored IDF arithmetic. This is an instance of the \emph{decision-determinism} stance the author has articulated separately [DetByDefault 2026]: consequential decisions (what fires, what routes, what abstains) computed by deterministic arithmetic over typed evidence, with the model running only where deterministic checks abstain. Here the stance is applied to the question of when a memory system may claim its reader's attention.

\subsection{The reliability layer: SRE for a memory subsystem}

The reliability layer is the paper's core proposal. It has three elements.

\textbf{The health gate.} At every session start, eleven checks probe the stack end to end and print one line, \texttt{MEMORY-HEALTH: 11/11 GREEN}, or name what is broken \emph{and how to fix it}. The checks cover ingest watermark freshness; vector coverage; per-hook heartbeats (did every hook that should have fired since the last boundary actually fire, and succeed?); capture-queue depth; lineage cache age; index and memory-file integrity; render novelty (a frozen session-start render across sessions is itself a failure); and an end-to-end embedding probe that \emph{discriminates} its failure modes (server unreachable vs. model missing vs. cold load vs. wedged) because each has a different fix. The design ancestor is the production-readiness rubric tradition in ML systems [Breck 2017] and the SRE monitoring canon [SRE 2016], transplanted to a memory subsystem. The gate is decision-deterministic end to end: every verdict is thresholded arithmetic over typed evidence: file ages against freshness bounds, heartbeat counts against expected hook sets, probe outcomes against an enumerated failure taxonomy, vector widths against a pinned dimension. No model participates in judging whether memory is healthy.

\textbf{Heartbeats.} Every hook invocation writes a success-or-failure row before doing its work. A hook that starts crashing therefore shows up as failure rows, and a hook that stops firing entirely shows up as an absence the heartbeat check converts to a RED at the next session boundary, including the gate's own absence, since the gate leaves heartbeats too. The design goal is that \textbf{no enumerated failure mode can pass unrecorded}. The enumeration has known residuals, which we state rather than hide: a crash before the heartbeat write itself, failure of the telemetry store, and the window between boundaries (a mid-session death is recorded but not \emph{flagged} until the next gate run). Within the enumerated modes, the record is clean: all 85 hook failures in the instrumented lifetime were recorded (\S4.1); none was discovered by suspicion.

\textbf{Alert economics.} The gate's contract borrows the SRE maxim that every page must be actionable [SRE 2016], and clinical alarm-fatigue research supplies the failure mode when it is violated: operators desensitize and override [DrugAlerts 2006; Ancker 2017]. Three consequences are built in. A \emph{cold} embedding server is not an outage: on probe timeout the gate takes one generous warm-up retry and, if the model comes up, reports GREEN with the warm-up named, instead of paging the operator to "retry." An \emph{optional} integration that was never wired reports green-with-a-note, not RED; a fresh install must not carry a permanent alarm for a tool it does not use. And \emph{misconfiguration is caught at configuration time}: a fallback embedding model whose vector width can never match the index is detected from model metadata at the gate, not discovered as a mysterious latency tax on failure paths months later (\S4.3, incident 2). Under this contract a RED is actionable by definition; the one RED class that expects a specific later self-heal (a capture event recorded outside any session boundary) states that expectation as its action ("verify green at the next boundary"), and its self-heal has been verified in production (\S4.2).

The gate also watches what the harness does not own: the project pairs with an external code-graph tool for structural queries, and a gate check monitors the extracted corpus's freshness; a stale graph silently answering dependency questions is the same failure class as a dead vector leg.

\subsection{Conventions as shipped artifacts}

The last component is not code. Some continuity only survives compaction if it is \emph{an artifact the workflow forcibly revisits}, not a habit and not merely context. Recent evidence sharpens the point: compaction can silently erase standing safety constraints from context over long horizons [GovDecay 2026], and vendor guidance now prescribes cross-session artifacts precisely because "compaction isn't sufficient" [Anthropic 2025b]. The harness therefore ships, as templates: the one-line-per-memory index with its resume anchor (read first after every compaction); invariant files that the constitution requires \emph{actively re-reading} at every arc boundary, on the theory that passively injected context gets skimmed; the append-only per-arc ledger (append at every decision, never rewrite, read the tail first on resume); and a constitution of standing rules.

One convention deserves emphasis because it is the policy answer to the two-directions problem (\S5, principle 4): the \textbf{tiered ground-truth hierarchy}. Running code and live terminal output outrank facts freshly derived this session, which outrank narrative documents and memory, which outrank the model's training knowledge. The tiers carry opposed obligations: tier-2 facts must be \emph{used without re-derivation} (anti-rediscovery), while tier-3 claims must be \emph{verified before acting}: a named symbol from a months-old document gets grepped before anything is built on it (anti-decay). Empirical study of exactly this artifact genre (agent-facing context files) has just begun [AgentREADME 2026]; our contribution is the pairing of the artifact with an arbitration rule between its failure directions.

\section{The operational record}

All numbers derive from the instrument stores by scripted queries, each documented in the artifact's numbers appendix. The related-work sweep behind \S8 was web-verified on 2026-08-30, and the three most load-bearing comparisons were additionally re-verified against live sources at drafting time.

\subsection{The long haul: 78,933 invocations and a maturation curve}

\begin{figure}[t]\centering
\includegraphics[width=\linewidth]{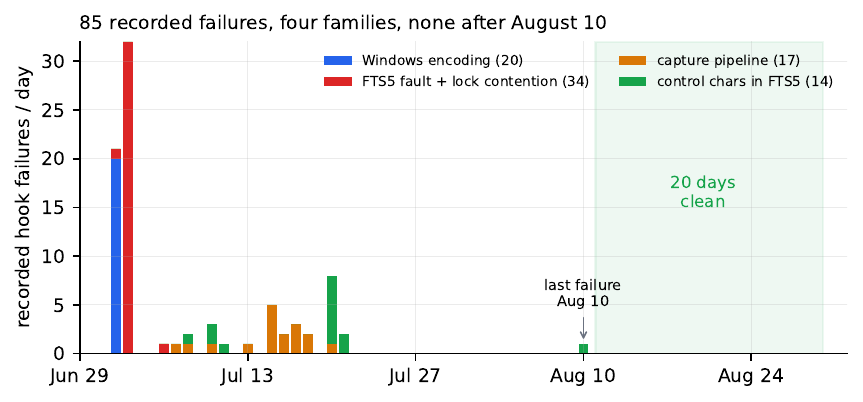}
\caption{All 85 recorded hook failures by day and family, classified from each row's stored error detail. Structural fixes ended each family, except four July 15--17 duplicate-chunk collisions that stopped without an attributable fix; the last recorded failure is 2026-08-10.}\label{fig:failures}
\end{figure}

From the reliability layer's first day (2026-07-02) through 2026-08-30, the heartbeat store records \textbf{78,933 hook invocations}. Eighty-five failed: \textbf{84 of them in the subsystem's first three weeks (July 2--21), one since (August 10), none in the final 20 days}. The failures cluster into four families, and the family structure \emph{is} the maturation story:

\begin{itemize}\itemsep2pt
\item \textbf{Day one, 2026-07-02:} 20 Windows text-encoding failures across four hooks: the platform's default code page rejecting real transcript content. Fixed by enforcing UTF-8 discipline at every entry point (now pinned by tests in the released kit). Day one also logged the first row of the next family, hours ahead of its burst.
\item \textbf{2026-07-02 $\to$ 07-06:} 34 failures: an FTS5 virtual-table fault plus database-lock contention under concurrent hooks. Root-caused to an unbounded FTS5 query pattern; bounded, and never recurred.
\item \textbf{July 7--20:} 17 capture-pipeline failures in two modes. Thirteen were aborts from writes to a dead standard-output pipe: the spawning process can exit before a long SessionEnd body finishes, the hook's informational echo then raises OSError(22), and the raise was aborting the \emph{remaining} ingest work. The fix absorbs hook-echo write failures (when the reader is gone the message has no destination; the ingest must still run). Four were duplicate-chunk collisions under concurrent capture passes (July 15--17); they did not recur, and no fix is attributable; we record those four as luck, not maturation.
\item \textbf{July 8 $\to$ August 10:} 14 low-rate failures from control characters in ingested text breaking FTS5 string parsing ("unterminated string"), ending with input sanitization at the boundaries, the last recorded hook failure of any kind.
\end{itemize}

Two readings. As a reliability record: a new memory subsystem in real production converges from a double-digit-failure first week to \textbf{a single failure in its last five weeks and none in its last 20 days}, and the trajectory is visible only because the heartbeat rows exist. As a claim about silence: we do not claim the system did not fail; we claim that within its enumerated failure modes it \textbf{could not fail without leaving a record}, and that not one of the 85 recorded failures was discovered by suspicion or archaeology.

\subsection{The instrumented window: ten days at full resolution}

\begin{figure}[t]\centering
\includegraphics[width=\linewidth]{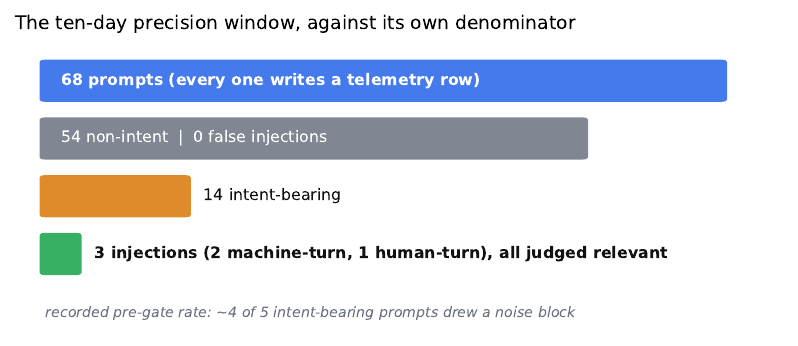}
\caption{The precision window against its own denominator: 68 prompts, 14 intent-bearing, 3 injections (all judged relevant), 0 false fires on the 54 non-intent prompts.}\label{fig:window}
\end{figure}

The 2026-08-19 $\to$ 08-28 window (ten calendar days, bounded by the precision instrument's deployment date on one side and this draft on the other) adds denominator-complete injection telemetry. It was a light-duty period by the project's standards (2,121 hook invocations, roughly a sixth of the lifetime daily average, spanning a multi-day idle gap and a network outage), which bounds what it can show about load, though not about precision.

\textbf{Turn taxonomy, defined once:} a \emph{prompt} is any UserPromptSubmit event (68 in the window; every one writes a telemetry row). Prompts divide into \emph{machine-generated} turns (agent-completion callbacks and similar mechanical events) and \emph{human-authored} turns; independently, the intent detector marks 14 prompts \emph{intent-bearing} (build/investigate/decide). Five injection firings on the deployment day were authors' synthetic validation probes and are excluded from use-analysis by that stated rule; they appear in the appendix.

\begin{table}[t]\centering\small
\begin{tabular}{p{0.30\linewidth}p{0.62\linewidth}}
\toprule
Measure & Value \\
\midrule
Hook invocations / failures & 2,121 / 0 \\
Prompts / intent-bearing & 68 / 14 \\
Injections on real-use turns & 3, each topically justified (author-adjudicated; \S6); 2 landed on machine-generated turns, motivating the system-turn exemption \\
False injections on non-intent turns & 0 of 54 \\
Retrieval latency, all prompts & median 0 ms (54 of 68 prompts trigger no retrieval work) \\
Retrieval latency, when retrieval ran (n=14) & median 2.5 s \textperiodcentered{} max 10.3 s (the pre-fix failure-path day; incident 2) \\
Gate verdicts & every RED genuine (cold server; aged corpus; stale watermark; one designed boundary-deferred capture check), each with a correct fix hint; zero false alarms; the deferred check's self-heal verified at the 08-28 boundary \\
\bottomrule\end{tabular}
\caption{The ten-day precision window (2026-08-19 to 08-28).}
\label{tab:window}\end{table}

Under the old pre-gate behavior (the stale-claim block firing on roughly four in five intent-bearing prompts), a 14-intent-prompt window would have produced on the order of eleven noise firings. The instrumented window produced three, all judged relevant, and zero on the 54 non-intent prompts. We present the old rate as recorded context (its sample denominator was not preserved); the window's own denominator is complete, and the adjudication of "topically justified" is the authors', a construct-validity limit stated in \S6.

\subsection{Three incidents and a delivery case}

Each incident: symptom $\to$ the instrument that caught it $\to$ root cause $\to$ structural fix.

\textbf{Incident 1: the day-one latency stack (11.3 s hook wall time $\to$ 1.6 s).} On the precision instrument's first day, the prompt hook's wall time exceeded its 3-second budget nearly fourfold, visible immediately in the invocation telemetry. Two stacked root causes: the client resolved \texttt{localhost} to IPv6 first while the local inference server binds IPv4-only, burning \textasciitilde{}2 seconds per call on a blackholed connect (the server was answering in 57 ms); and the gate's warm-up probe carried no context-size option while the real embed path sends one, so the model server \emph{reloaded the model twice per hook invocation} to satisfy the mismatched context lengths. Structural fixes: literal IPv4 loopback defaults everywhere (with the rationale in a comment), and probe/request option parity. Lesson: the kit's 580 offline tests did not catch either root cause and, being fully offline, could not have; the production instrument is the test rig for live-latency defects.

\textbf{Incident 2: the impossible fallback and the failure-path chain.} Degradation telemetry showed cold-start windows costing far more than a cold model should. Trace: a configured \emph{fallback} embedding model emitted 768-dimensional vectors against a 1024-dimensional index: the dimension guard refused it every time, so the fallback could never succeed; it existed only as an \textasciitilde{}18-second model load plus a second timeout chain appended to every failure path. Worst case, one prompt's injections were computed and then lost at 8.1 seconds against the 3-second hook budget. Structural fixes: a per-process circuit breaker (first embed failure opens the circuit; sibling searches within the doomed hook bypass the retry; the next short-lived hook process retries fresh), a widened budget, and, the general lesson, \textbf{configuration validated at configuration time}: the gate now reads the fallback model's dimension from server metadata and REDs an impossible fallback the day it is configured.

\textbf{Incident 3: injections nobody reads.} The precision telemetry's first full-window read showed two of the three real-use injections had fired on \emph{machine-generated} turns: completion callbacks that flow through the same prompt hook as human input and present no decision point to any reader, human or agent. Topical relevance was genuine; readership was zero; the spend was real. Fix: the system-turn exemption (\S3.3), with the telemetry row still written so the denominator survives. Lesson: relevance is not sufficient; the injection contract is with a reader, and a turn can have none.

\textbf{The delivery case.} The window's remaining real-use injection fired on a human-authored intent-bearing turn: DO-NOT-REBUILD, stale-claim, and lineage blocks cleared every gate and named existing production subsystems relevant to the discussion at hand. We record this as what the telemetry shows, \emph{delivery}: the right record surfaced at the moment of intent. Whether delivery becomes \emph{prevention} (work not repeated that otherwise would have been) is precisely the counterfactual an observational record cannot count (the pre-harness contrast is \S3.2's three-session re-proposal incident), and it is the primary quantity the pre-registered A1 ablation (\S7) is designed to measure.

\section{Design principles}

Several of these principles are instances of a single stance carried forward from the author's prior work, \textbf{decision-determinism} [DetByDefault 2026]: every consequential decision in the memory stack (what fires, what REDs, what is refused at the door) is computed by deterministic code over typed evidence, and the model is never the judge of its own memory's health or relevance.

\begin{enumerate}\itemsep2pt
\item \textbf{Memory is infrastructure; run it like infrastructure.} Give the memory subsystem the treatment the serving stack gets: health checks with discriminated failure modes, heartbeats, freshness objectives, and a single legible verdict at every session start.
\item \textbf{Instrument the denominator, not just the hits.} Every hook invocation writes a row \emph{before} doing work, including the boring ones. Precision claims, failure rates, and "did it stop firing?" are then queries, not recollections.
\item \textbf{Precision beats recall for injected context.} The binding constraint is the reader's trust in the channel, spent by every irrelevant firing and only slowly re-earned. Silence must be the default; a retrieval layer the reader has learned to skim is worse than none.
\item \textbf{Fight both failure directions at once, with opposed rules.} Memory fails toward \emph{rediscovery} (what exists is ignored) and toward \emph{decay} (what is recorded is wrong). The mitigations pull in opposite directions ("use it without re-deriving" versus "verify before acting") and must be held in tension by an explicit arbitration rule (our tiered ground-truth hierarchy), not averaged into mush.
\item \textbf{Alert economics are load-bearing.} Every RED must be actionable; cold is not dead; optional is not broken; misconfiguration should fail at configuration time. An unactionable alarm does not merely waste attention; it trains the reader out of the channel that will one day carry the real one.
\item \textbf{Continuity must be an artifact, not a habit.} Anything that must survive compaction gets a file the workflow forcibly revisits: resume anchors read first, invariants re-read at boundaries, append-only ledgers whose tails are read on resume. Context injection is a delivery mechanism, not a guarantee of uptake.
\item \textbf{Boring storage, honest degradation.} Spend novelty on contracts (refuse mismatched vectors; degrade search rather than block; log every degradation once) rather than on representation. Every component sophisticated enough to fail interestingly will.
\end{enumerate}

\section{Limitations and threats to validity}

This is an experience report from a single deployment, and its evidential limits are structural. \textbf{N = 1}: one project, one human operator with one agent, one stack (Claude Code, local Ollama, Windows), one domain. \textbf{No control arm}: the project was never run without the harness over a comparable period; effect sizes are not causal claims. \textbf{Confounding}: the eight months span model generations, tooling improvements, and operator learning, all of which improve outcomes independently of memory infrastructure. \textbf{Construct validity}: the labels doing evidential work ("topically justified," "intent-bearing," "false fire") are author-adjudicated with no independent rater; the pre-gate baseline's sample denominator was not preserved and is reported as context only. \textbf{Window representativeness}: the precision window is bounded by its instrument's deployment date rather than chosen, but it was a light-duty period (about a sixth of lifetime average load) including an idle gap and an outage. \textbf{Observer effects}: a shakedown run by the system's builders changes the builders' behavior. \textbf{Self-report}: the record is produced by the system's own operator-builders, and one of those operators is the agent the system serves; the instrumentation is designed to resist motivated reading (denominators, append-only stores), but design intent is not independence. \textbf{Unmeasured costs}: we did not instrument the harness's own overhead: triage minutes, injected tokens, maintenance attention; a net-cost accounting is required before claims of net benefit, and the counter-position discussed in \S8 is fundamentally a net-cost thesis. \textbf{Generality of the principles vs. the record}: the design stance behind \S5 is not unique to this deployment: the same operator pair applied it to a second, independently published system, a metadata-reconstruction verification harness [Rosetta 2026] whose companion methodology paper [DetByDefault 2026] articulates the shared stance; that second instantiation tempers the single-deployment concern for the \emph{principles}, while the memory-subsystem operational record itself remains N=1. \textbf{Scope}: a single-operator, local-only deployment sidesteps threats that matter at team scale, most notably memory poisoning, which can persist across sessions and activate much later [AgentPoison 2024; MINJA 2025]; a shared-memory deployment would need admission control and provenance we did not need. Finally, the quantities we care most about (rediscoveries \emph{prevented}, knowledge \emph{not} lost) are counterfactuals observation cannot count. That is what \S7 is for.

\section{A pre-registered ablation protocol}

The released kit makes the causal questions testable, on our deployment or anyone's. The full protocol is committed and tagged in the artifact repository (\texttt{PREREGISTRATION.md}, tag \texttt{prereg-ablation-v1}, 2026-08-30, registered before any ablation run); we summarize it here.

\begin{itemize}\itemsep2pt
\item \textbf{A1: injection off, 14 days.} Emission disabled; intent detection, relevance gating, and telemetry stay on, so every suppressed injection is recorded as a \emph{would-have-fired} row and the denominator survives. Primary metric: \emph{re-proposal incidents}, mechanically defined: a build-intent prompt whose would-have-fired DO-NOT-REBUILD record names an existing subsystem \emph{and} whose session subsequently begins an overlapping new implementation; the overlap judgment is scored independently by both operators (human and agent), blind to each other, with disagreements reported rather than resolved. Delivery alone is never counted as prevention. Comparison: the preceding 14 days, as rates per intent-bearing prompt (workload varies several-fold). Decision rule: supported if the ablated window shows $\geq$2 scored incidents and the intact window fewer at comparable volume; refuted on this axis if 0 incidents accrue against $\geq$5 would-have-fired records; otherwise \emph{inconclusive, reported as such}. Latency and interruption counts are reported but not creditable to the ablated arm; disabling emission reduces both by construction.
\item \textbf{A2: gate off, 14 days.} The gate is not run; heartbeats and degradation logs still record. Primary metric: time from a naturally occurring degradation's first evidence row to first remediation, against the intact configuration's boundary-catch record. Supported if $\geq$2 degradations occur with median detection latency $\geq$4$\times$ the intact median; inconclusive below 2 occurrences (the record's base rate, roughly one degradation per multi-day idle gap, makes a quiet fortnight possible, and a quiet fortnight proves nothing).
\end{itemize}

We commit to reporting results regardless of direction, and we invite hostile replication: the kit installs on any Claude Code project in one command, and every metric above is a query against stores the kit creates.

\section{Related work}

\noindent\emph{Sweep date: 2026-08-30. This space is currently minting close neighbors monthly; the sweep should be re-run at any later submission.}

Three research streams converge on, and stop short of, this paper's ground.

\textbf{Memory architectures.} From MemGPT's OS metaphor [MemGPT 2023] and the generative-agents memory stream [Park 2023] through production-marketed systems (Mem0's extraction pipeline [Mem0 2025], Zep's bi-temporal knowledge graph [Zep 2025], A-MEM's linked notes [AMEM 2025], HippoRAG's graph retrieval [HippoRAG 2024]), the architecture axis is rich and benchmark-driven, and MemOS goes furthest toward our framing by declaring memory "a first-class operational resource" with governance abstractions [MemOS 2025]. Surveys organize the space by mechanism and capability [CoALA 2024; TOIS-survey 2025; AgentMemSurvey 2025]. The agent-capability corollary (that harness and interface, not just model, determine outcomes) is the ACI thesis [SWE-agent 2024], of which this paper is the memory-subsystem instance at months scale. What none of this line reports is the \emph{operation} of a memory system in production: the Always-On Agents survey, coding 435 works, finds the literature "concentrates more heavily on accumulating and retrieving state than on governing, recovering, or relinquishing it" [AlwaysOn 2026], a survey-scale statement of the gap this paper fills from the field side, and one the ML-systems tradition would recognize as data-dependency debt wearing a new face [Sculley 2015].

\textbf{Memory failure studies (the 2026 wave).} A rapid recent line argues deployed agent memory degrades and must be engineered for: aging mechanisms and lifespan evaluation [Aging 2026], per-operation failure taxonomies under stress [MemFail 2026], staleness and validity benchmarks [STALE 2026; ATMA 2026; TemporalValidity 2026], and a dual metric penalizing both recall failure and reuse of invalidated memory [FAMA 2026]. We regard this wave as the in-vitro companion to our in-vivo record: it names and measures, under controlled faults, the failure families our incidents exhibit under natural ones. On the practice side, SRE ideas have been applied to agents \emph{in general} (safety SLOs, circuit breakers, health checks [MsftSRE 2026, as described in its published summary]) and agent-observability tooling now traces memory operations as spans [Braintrust 2026]; we found no prior report of an operated, health-gated, alert-budgeted reliability layer for a \emph{memory subsystem specifically}, which is the practice this paper documents. Adjacent from the verification side is the author's own prior line: placing a language model inside a deterministic verification harness, with provenance-carrying claims and evidence-class-bounded confidence, demonstrated on metadata reconstruction [Rosetta 2026] and distilled as design principles [DetByDefault 2026]. There the harness governs the \emph{model's claims about data}; here the same stance governs the \emph{memory subsystem of a development agent}.

\textbf{Production experience with AI-assisted development.} The genre runs from population-scale telemetry [Ziegler 2024; Peng 2023] through org-scale deployment reports (25 months of GenAI telemetry at WhatsApp [WhatsCode 2026], LLM-driven migrations at Google [GoogleMigrate 2025]) to RCTs whose disagreement about productivity [METR-RCT 2025] underlines that throughput is not the only quantity worth measuring. These reports are longitudinal but memory-blind. The nearest neighbor to our record is \emph{Written by AI, Managed by AI} [WrittenByAI 2026]: 391 consecutive LLM sessions on one real project over about a month, qualitative action research that names an "Index Sickness" failure pattern and concludes that \emph{more} formal constraint scaffolding (symbolic identifier systems, accumulated defensive rules in standing prompts, longer contexts) degraded long-project performance, with a minimal baseline-log fix. The regimes differ: their indicted scaffolding is \emph{standing symbolic constraint}, accumulating in every prompt, and their failure mechanism is agent-side symbolic retreat; our injection layer is \emph{episodic}, relevance-gated, silent by default, and health-monitored (their report describes no gating or monitoring layer), and our observed failure economics concern the reader's attention. The two records agree on more than they dispute: their successful fix is itself a compaction-surviving artifact of exactly \S3.5's kind, and both records testify that un-curated accumulation degrades the channel it fills. What our precision record shows is that the noise \emph{mechanism} is suppressible; whether gated injection yields net benefit where their un-gated scaffolding cost net performance is an open question our A1 protocol is designed to answer. Finally, the pre-AI software-engineering literature stated our problem first: knowledge loss under developer turnover, measured and priced [Rigby 2016; Robillard 2021]; documentation that goes stale silently [OutdatedDocs 2023]; domain knowledge thinly spread as a root cause of failure [Curtis 1988]; and developers relying on interruption and archaeology because knowledge lives in heads [LaToza 2006]. A months-long agent session line is, in effect, a developer who turns over at every compaction; that is why turnover economics, not context-window engineering, is the right ancestral frame.

\section{Conclusion}

At task scale, agent memory is a convenience. At tenure scale it is load-bearing infrastructure, and it fails like infrastructure: quietly, at the joints, in both directions at once. The record reported here (78,933 invocations; 85 recorded failures, none silent, none in the final 20 days; a noise channel remade so that silence is the default and every firing is accountable to a denominator; three incidents traced from instrument to structural fix) is one deployment's evidence that the reliability-engineering toolkit transplants cleanly to agent memory (and transplants decision-deterministically, the model never adjudicating its own continuity), and that the questions it cannot settle can at least be made precise, pre-registered, and cheap for others to run.

\section*{Artifact availability}

SIx Harness v0.2.2 is released under the MIT license at \texttt{github.com/mike-m6online/SIx\_Harness} (an archival DOI deposit accompanies submission). The repository includes the 580-test offline suite, the one-command installer, and the tagged pre-registration (\texttt{prereg-ablation-v1}). Every number in this paper is re-derivable from the documented queries in the paper's numbers appendix run against our instrument stores; the stores themselves contain project-identifying content and are not released; the queries, schemas, and the instrument code that produces such stores on any deployment are. The name "SIx" is capitalized as in the origin system's convention, not a typographical error.

\section*{Acknowledgments}

This system was co-built, and is daily co-operated, with Claude (Anthropic), through Claude Code, the environment the harness serves. The agent implemented much of the infrastructure under the author's direction, operates its triage and health routines, and assisted substantially in drafting this manuscript, including deriving every reported number from the instrument stores. Per arXiv policy, AI systems are not listed as authors; the author takes full responsibility for all claims.

\end{document}